\documentclass[twocolumn,tighten]{aastex631}

\usepackage{amsmath}
\definecolor{mygreen}{rgb}{0.19,0.55,0.11}

\shorttitle{Modeling the evolution of CO+He WD binaries}
\shortauthors{Chen et al.}

\begin{document}

\title{Evolution of AM CVn binaries with WD donors}

\correspondingauthor{Hai-Liang Chen}
\email{chenhl@ynao.ac.cn}

\author{Hai-Liang Chen}
\affiliation{Yunnan Observatories, Chinese Academy of Sciences (CAS), Kunming 650216, P.R. China}

\author{Xuefei Chen}
\affiliation{Yunnan Observatories, Chinese Academy of Sciences (CAS), Kunming 650216, P.R. China}
\affiliation{University of the Chinese Academy of Sciences, Yuquan Road 19, Shijingshan Block, 100049, Beijing, China}

\author[0000-0001-9204-7778]{Zhanwen Han}
\affiliation{Yunnan Observatories, Chinese Academy of Sciences (CAS), Kunming 650216, P.R. China}
\affiliation{University of the Chinese Academy of Sciences, Yuquan Road 19, Shijingshan Block, 100049, Beijing, China}

\begin{abstract}

The evolution and stability of mass transfer of CO+He WD binaries are not well understood. Observationally they may emerge as AM CVn binaries and are important gravitational wave (GW) emitters. 
In this work, we have modeled the evolution of double WD binaries with accretor masses of $0.50 - 1.30\;M_{\odot}$ and donor masses of $0.17\; - 0.45\;M_{\odot}$ using the detailed stellar evolution code \textsc{mesa}. We find that the evolution of binaries with same donor masses but different accretor masses is very similar and binaries with same accretor masses but larger He donor masses have larger maximum mass transfer rates and smaller minimum orbital periods.  We also demonstrate that the GW signal from AM CVn binaries can be detected by space-borne GW observatories, such as LISA, TianQin. And there is a linear relation between the donor mass and gravitational wave frequency during mass transfer phase. In our calculation, all binaries can have dynamically stable mass transfer, which is very different from previous studies. The threshold donor mass of Eddington-limited mass transfer for a given accretor WD mass is lower than previous studies. Assuming that a binary may enter common envelope if the mass transfer rate exceeds the
maximum stable burning rate of He, we provide a new criterion for double WDs surviving mass transfer, which is below the threshold of Eddington-limit. Finally, we find that some systems with ONe WDs in our calculation may evolve into detached binaries consisting of neutron stars (NSs) and extremely low mass He WDs and further ultra-compact X-ray binaries.

\end{abstract}

\keywords{: Close binary stars (254); White dwarf stars(1799); AM Canum Venaticorum stars (31); Compact binary stars (283); Gravitational wave sources (677))}

\section{Introduction} 
\label{sec:intro}

AM CVn binaries are a kind of interacting binary systems consisting of accreting white dwarfs (WDs) and He-rich donor stars. They are important for studies of binary evolution \citep[e.g.][]{py14,tv23}, binary population synthesis \citep[e.g.][]{npvy01,hgcc20} and common envelope evolution \citep[e.g.][]{ijcd+13,kndg+21}. Given the short periods ($\sim 5 - 66\;$min) of AM CVn binaries, they are important gravitational wave (GW) sources for  space-borne low-frequency GW observatories like LISA \citep{aabb+17}, TianQin \citep{lcdg+16}, and Taiji \citep{rgcz20}. It was also suggested that AM CVn binaries can be the progenitors of type Ia supernovae \citep[][]{bswn07}. 

It is known that there are three possible formation channels for AM CVn binaries. In the first channel, a WD accretes material from a semidegenerate He star \citep[e.g.][]{ty79,nrs81,it91,yung08}; In the second channel, a He WD in a double WD system transfers material to the WD accretor \citep[e.g.][]{nrs81,ty96,npvy01,ctch22a}. In this channel, the accretor is usually a CO WD. In the third channel, an evolved main sequence donor star starts mass transfer around the end of main sequence. After the donor star loses its H-rich envelope, it becomes He-rich and has a remaining mass smaller than $0.10\;M_{\odot}$, transferring material to the WD \citep[e.g.][]{phr03,ljc21}. In this paper, we mainly focus on the double WD channel. 

Regarding the evolution of AM CVn binaries in the double WD channel, it has been widely investigated. Some studies \citep[e.g.][]{npvy01,mns04,gpf07,kblk17} adopted a semi-analytic method to model the evolution of AM CVn binaries. In these studies, the detailed structure of the He WD was not taken into account and the He WD was assumed to be fully degenerate. This should not be realistic since the He WDs may have small but thick envelopes. \citet{kbs12} have shown that this will have an important impact on the evolution of WD binaries and stability of mass transfer. In addition, some studies \citep[e.g.][]{dtwc07,wb21} have modelled the evolution of AM CVn stars with the He WD structure considered. They found that the initial entropy of the He WD has an impact on the evolution of AM CVn binaries. But it is worth noting that it is widely assumed that the mass transfer in AM CVn binaries is conservative and the evolution of accreted material on the CO WD is not considered in these studies. 

From previous studies of He-accreting WDs, we know that the evolution of He-accreting WDs strongly depends on the WD mass and accretion rate \citep[e.g.][]{nomo82b,it89,lt91,pty14,wph17,wwlh17}. 
It is shown that there is a stable burning regime in which the accreted He material can burn stably on the surface of WDs. There is little mass loss in this regime. If the accretion rate is smaller than the minimum stable burning rate, the He-burning is unstable, leading to nova outburst. A fraction of material can be lost during the nova outburst. If the accretion rate is larger than the maximum stable burning rate. The evolution is still under debate. \citet{py14} and \citet{wph17} found that the accreting WDs will evolve into red giants after a small amount of material is accreted. In this case, the binary system is likely to enter common envelope and merge eventually. On the other hand, \citet{hknu99} found that the optically thick wind \citep{kh94} will occur in this regime. In this scenario, the material on the surface of WDs burns at a rate of the maximum stable burning rate and the excess material is lost in a form of optically thick wind. With these results in mind, we can find that the mass transfer of AM CVn binaries is not likely to be conservative.

This work aims at a comprehensive study of the evolution of AM CVn binaries from double WD channel and the properties of this kind of binaries, in particular, as gravitational wave sources.  

The rest of the paper is organized as follows. In section~\ref{sec:met}, we describe how we make the initial WD models and briefly outline the assumptions underlying the binary evolution. In Section~\ref{sec:res}, we present the results we obtained. In Section~\ref{sec:dis}, we first discuss the detectability of AM CVn binaries with LISA, TianQin and the properties of AM CVn binaries as GW sources. Then we also discuss the stability of mass transfer of AM CVn binaries from our simulation. In addition, we discuss the uncertainties in our simulation and their influence on our results. Finally, we summarize our conclusion in Section~\ref{sec:con}.

\section{Method and assumptions} 
\label{sec:met}

\subsection{Initial He WD models}
\label{sec:ini_wd}

In this work, we adopted the He WD models from our previous work about the evolution of NS+He binaries \citep{ctch22b}. Here we briefly describe how these He WD models are made. 

First, the evolution of a grid of low-mass X-ray binaries was computed with the stellar evolution code \textsc{mesa} \citep{pbdh+11,pcab+13,pmsb+15,psbb+18,pssg+19}. The initial NS and donors are assumed to be a point mass and zero-age main-sequence stars, respectively. Their masses are assumed to be $1.30\;$ and $1.20\;M_{\odot}$, respectively. 
The initial orbital periods range from $1.0$ to $600\;$days. In this grid, some donors can evolve into He WDs. Then we extract the WD models when the central temperatures of He WDs are around $10^{7}\;$K.  In these models, the He WDs have H envelopes with masses smaller than $0.01\;M_{\odot}$.
The initial He WD masses are $0.17$, $0.21$, $0.25$, $0.30$, $0.35$, $0.40$, $0.45\;M_{\odot}$. Compared with the He WD models in \citet{lcch19} who modeled the formation of CO+He WDs, our He WD models may have different temperatures and H envelope masses. But from the following discussion, we can find these factors does not influence our results significantly. 

\subsection{Binary evolutionary models}

To model the evolution of AM CVn binaries, we make use of the \textsf{star\_plus\_point\_mass} test suite of \textsc{mesa} code (version 12115). In our simulation, the WD accretor is assumed to be a point mass. The initial masses of WD accretors are assumed to be $0.50$, $0.60$, $0.70$, $0.80$, $0.90$, $1.00$, $1.10$, $1.20$, $1.30\;M_{\odot}$. These WDs with masses larger than $1.10\;M_{\odot}$ are ONe WDs and others are CO WDs. 
The initial orbital periods are assumed to be $0.05\;$days. 

In our calculation, we consider two kinds of mechanisms of angular momentum loss: GW radiation and angular momentum loss due to mass loss. The angular momentum loss due to GW radiation can be computed with the following formula \citep{ll71}:
\begin{equation}
	\frac{{\rm d}J_{\rm gw}}{{\rm d}t} = -\frac{32}{5}\frac{G^{7/2}}{c^{5}}\frac{M^2_{\rm a}M^2_{\rm d} (M_{\rm a}+M_{\rm d})^{1/2}}{a^{7/2}}\;,
\end{equation}
where $G$ is the gravitational constant and $c$ is the speed of light in vacuum; $a$ is the binary separation; $M_{\rm a}$ and $M_{\rm d}$ are the masses of the WD accretor and the WD donor, respectively. 

We compute the mass transfer rate with the \textsf{Ritter} scheme \citep{ritt88}. The mass transfer in our calculation is not conservative and the isotropic re-emission model are adopted \citep{tv06}. The mass retention efficiency is computed as follows. 

Given the H shell of the He WD, the accreted material is H-rich at the early phase of mass transfer. We simply assume that the retention efficiency is $0$ in this phase. We have tested that this assumption has little impact on our results. As for He burning, we adopt the optical thick wind model \citep{kh94,hkn99} and the prescription of \citet{kh04}. If the mass transfer rate ($\dot{M}_{\rm tr}$) is larger than a threshold value $\dot{M}_{\rm up}$, we assume that He burns steadily at a rate of $\dot{M}_{\rm up}$ and the excess material is lost in the form of optically thick wind. The threshold mass transfer rate is 
\begin{equation}
\dot{M}_{\mathrm{up}}=7.2 \times 10^{-6}\left(M_{\mathrm{a}} / \mathrm{M}_{\odot}-0.6\right)\;\mathrm{M}_{\odot}\;\mathrm{yr}^{-1},
\label{eq:up}
\end{equation}
for accretor masses $M_{\rm a} \ge 0.75\;M_{\odot}$ \citep{nomo82b}. 

If the mass transfer rate is larger than the minimum rate of stable burning ($\dot{M}_{\rm cr}$) and smaller than $\dot{M}_{\rm up}$, we assume that He burns stably on the surface of the WD and there is no mass loss. If the mass transfer rate is smaller than $\dot{M}_{\rm cr}$ but larger than $\dot{M}_{\rm low}$, we assume that the He burns unstably, triggering He flashes. In this regime, the retention efficiency is computed following \citet{kh04}. If the mass transfer rate is smaller than $\dot{M}_{\rm low}$, we assume that the He flashes is too strong to retain any material.  Therefore the retention efficiency for He burning is 

$$
\eta_{\mathrm{He}}=\left\{\begin{array}{ll}
\dot{M}_{\mathrm{up}} / \dot{M}_{\mathrm{tr}} & \dot{M}_{\mathrm{tr}}\ge \dot{M}_{\mathrm{up}} \\
1 & \dot{M}_{\mathrm{up}}>\dot{M}_{\mathrm{tr}} \ge \dot{M}_{\mathrm{cr}} \\
\eta_{\mathrm{KH04}} & \dot{M}_{\mathrm{cr}}>\dot{M}_{\mathrm{tr}} \ge \dot{M}_{\mathrm{low}} \\
0 & \dot{M}_{\mathrm{tr}}<\dot{M}_{\mathrm{low}}
\end{array}\right.
$$
where
$$
\begin{array}{l}
\dot{M}_{\mathrm{cr}}=10^{-5.8}\; M_{\odot}\; \mathrm{yr}^{-1} \\
\dot{M}_{\mathrm{low}}=10^{-7.4}\; M_{\odot}\; \mathrm{yr}^{-1}.
\end{array}
$$

For WDs with masses $M_{\rm a} < 0.75\;M_{\odot}$, Eq.~\ref{eq:up} is not validated any more and we simply assume that the mass transfer is completely non-conservative, i.e. no mass is retained by the accretor. These material not accreted by the accretor leaves the system and takes away the specific angular momentum of the accretor. 

In addition, we also take the Eddington limit into consideration. The Eddington limit can be given by \citep{tv23}
\begin{equation}
\dot{M}_{\rm Edd} = 4.4 \times 10^{-6}\;(M_{\rm a}/M_{\odot})\;M_{\odot}\;{\rm yr}^{-1}.
\label{eq:edd}
\end{equation}
Compared with the Eddington limit of \citet{hw99}, the Eddington limit in our calculation is lower. This can be understood as follows. 
In the calculation of Eddington limit, \citet{hw99} only considered the gravitational energy released by the accreted material. However, \citet{tv23} suggest that the nuclear burning energy from these accreted material should be also considered, which leads to a lower Eddington accretion rate. 
Following \citet{hw99}, we assume that the binary will merge in a common envelope if the mass transfer rate is larger than the Eddington limit. But we do not stop the calculation in order to know if the binary system can have dynamically stable mass transfer.  

The inlist files for our simulations can be made available on request by contacting the corresponding author.

\section{Results}
\label{sec:res}

\subsection{Examples of binary evolution}

\begin{figure}
    \centering
    \includegraphics[width=\columnwidth]{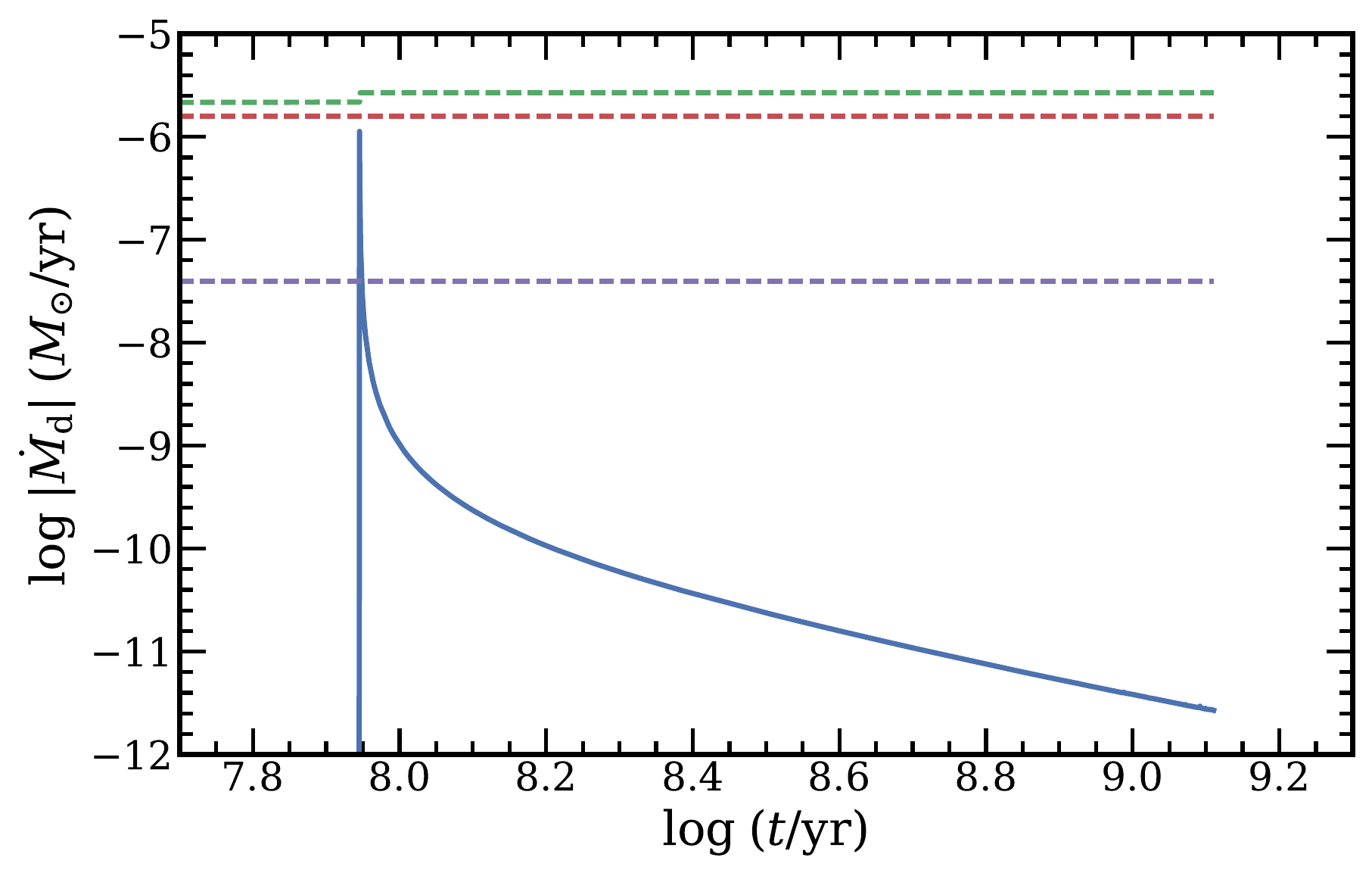}
    \includegraphics[width=1.1\columnwidth]{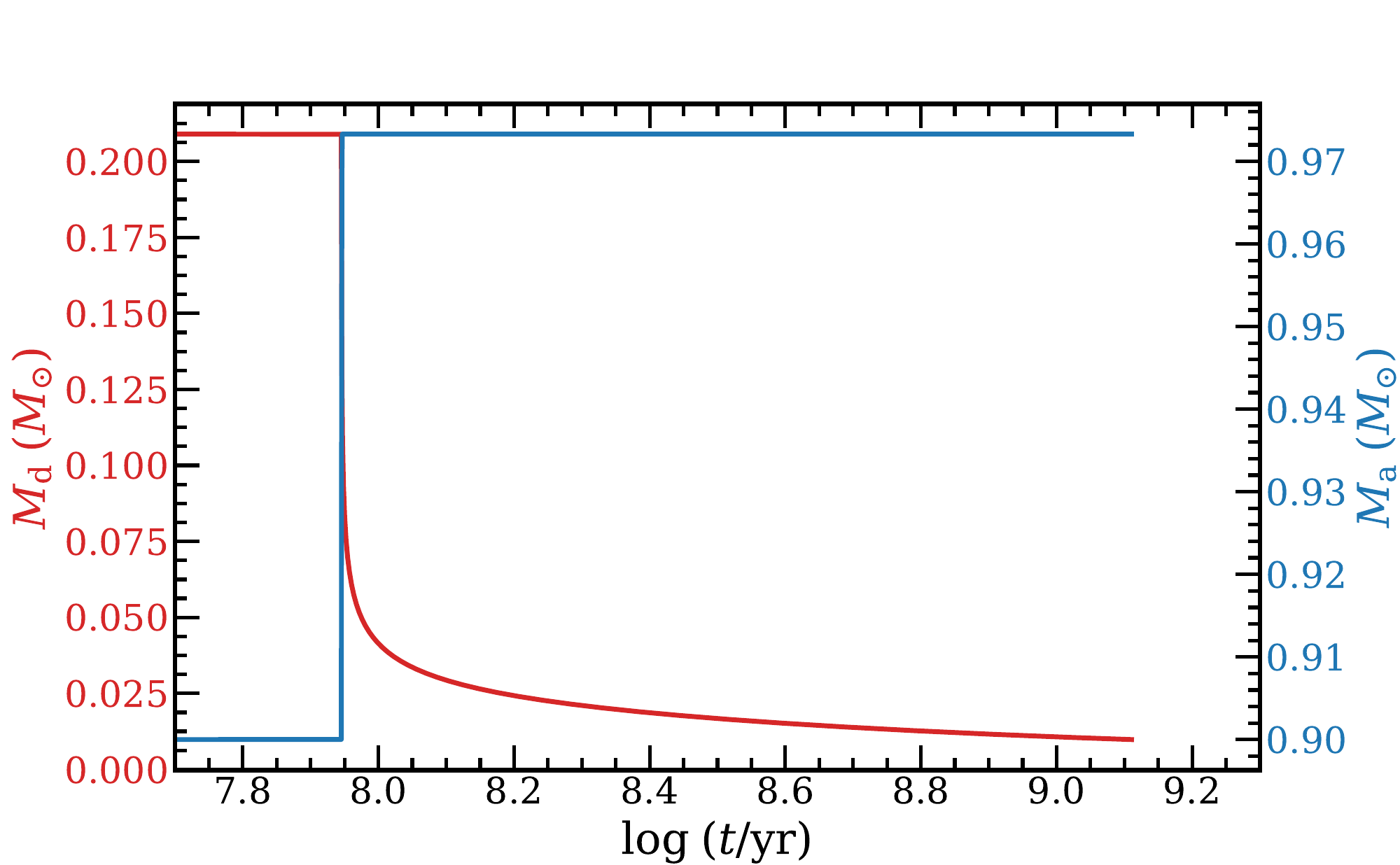}
    \includegraphics[width=\columnwidth]{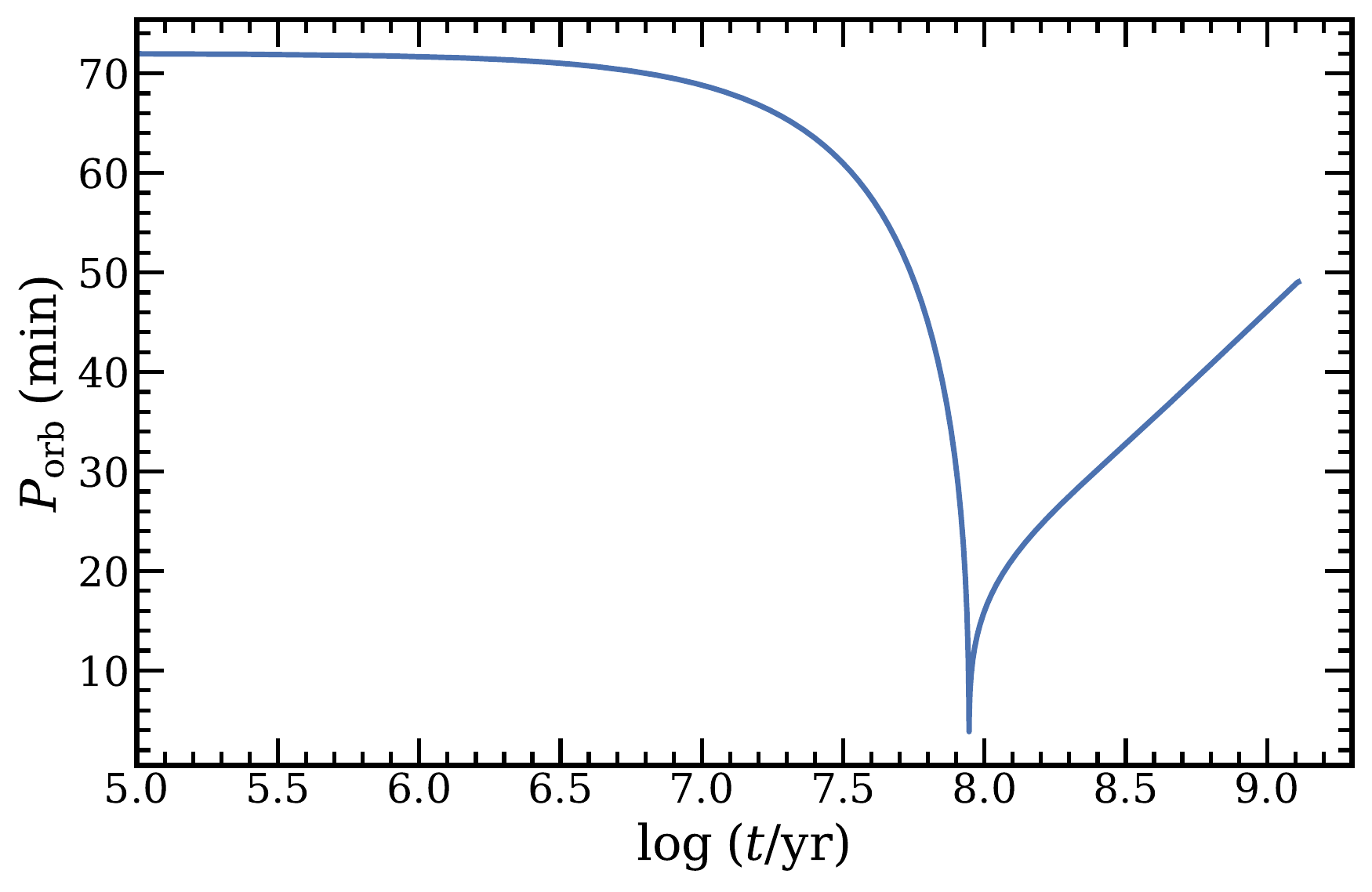}
    \caption{Evolution of mass transfer rate (upper panel), WD masses (middle panel) and orbital period (lower panel) as a function of time. The initial binary parameters in this example are $M_{\rm a} = 0.90\;M_{\odot}$, $M_{\rm d} = 0.21\;M_{\odot}$ and $P_{\rm orb} = 0.05\;$days. In the upper panel, the three dashed lines from up to bottom indicate $\dot{M}_{\rm up}$, $\dot{M}_{\rm cr}$ and $\dot{M}_{\rm low}$, respectively. In the middle panel, the red and blue lines are for the donor and accretor masses, respectively.    }
    \label{fig:sg_evl_ex}
\end{figure}

In Fig.~\ref{fig:sg_evl_ex}, we present an example of binary evolution of AM CVn binaries. The initial masses of the accretor and donor are $0.90\;M_{\odot}$ and $0.21\;M_{\odot}$, respectively. The initial orbital period is $0.05\;$days. From this plot, we can find that the mass transfer rate in this example is always below the stable burning regime of He (indicated by the green and red dashed lines in the upper panel). The mass of WD accretor increases when the mass transfer rate is between $\dot{M}_{\rm low}$ (see the purple dashed line) and $\dot{M}_{\rm cr}$ (see the red dashed line). At the early phase of evolution, the orbital period decreases because of GW radiation. When the H envelope is stripped, the mass transfer rate is around its maximum value and the orbital period reaches its minimum. Afterwards, the mass transfer leads to the increase of orbital period.                         

\begin{figure}
    \centering
    \includegraphics[width=\columnwidth]{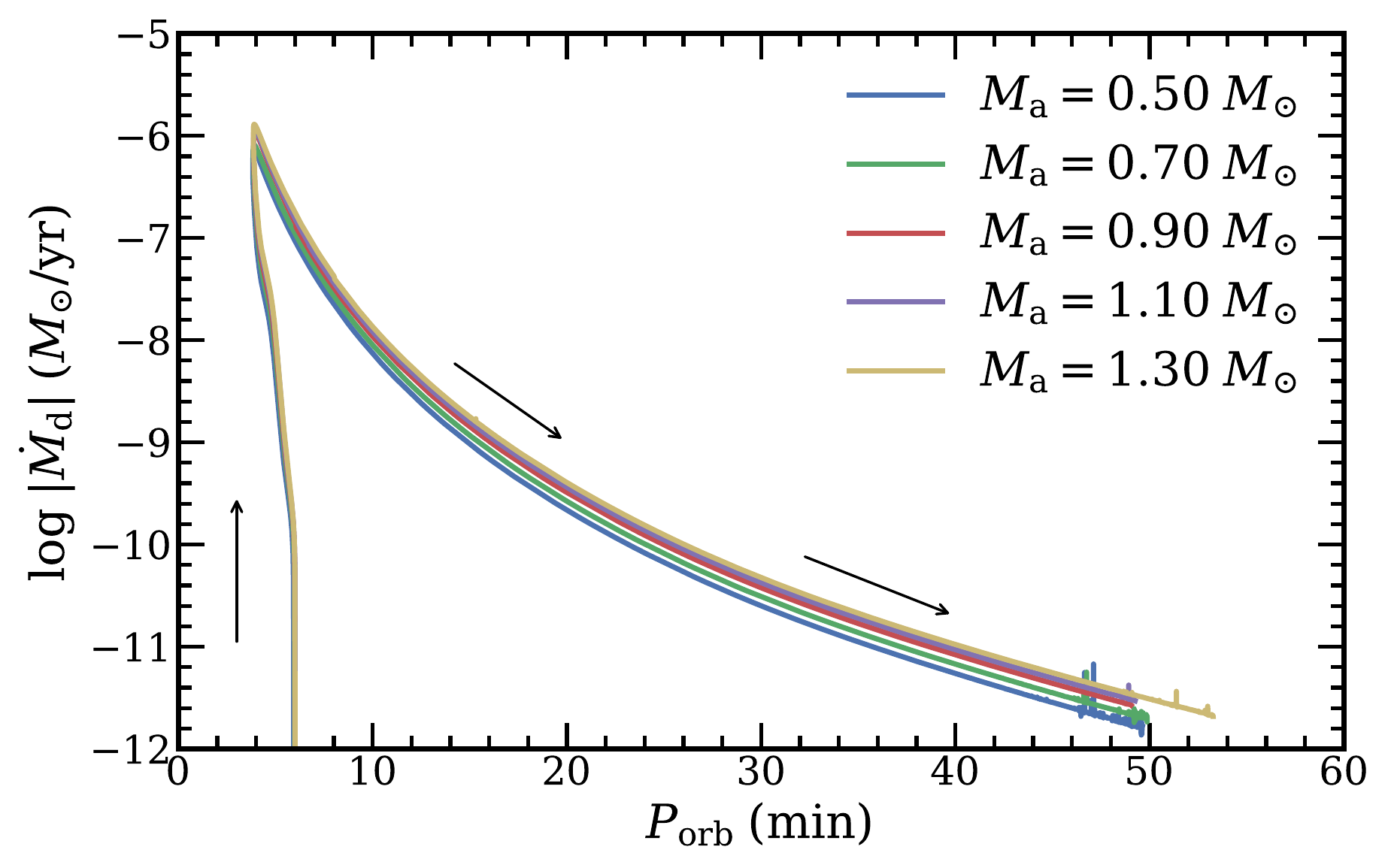}
    \includegraphics[width=\columnwidth]{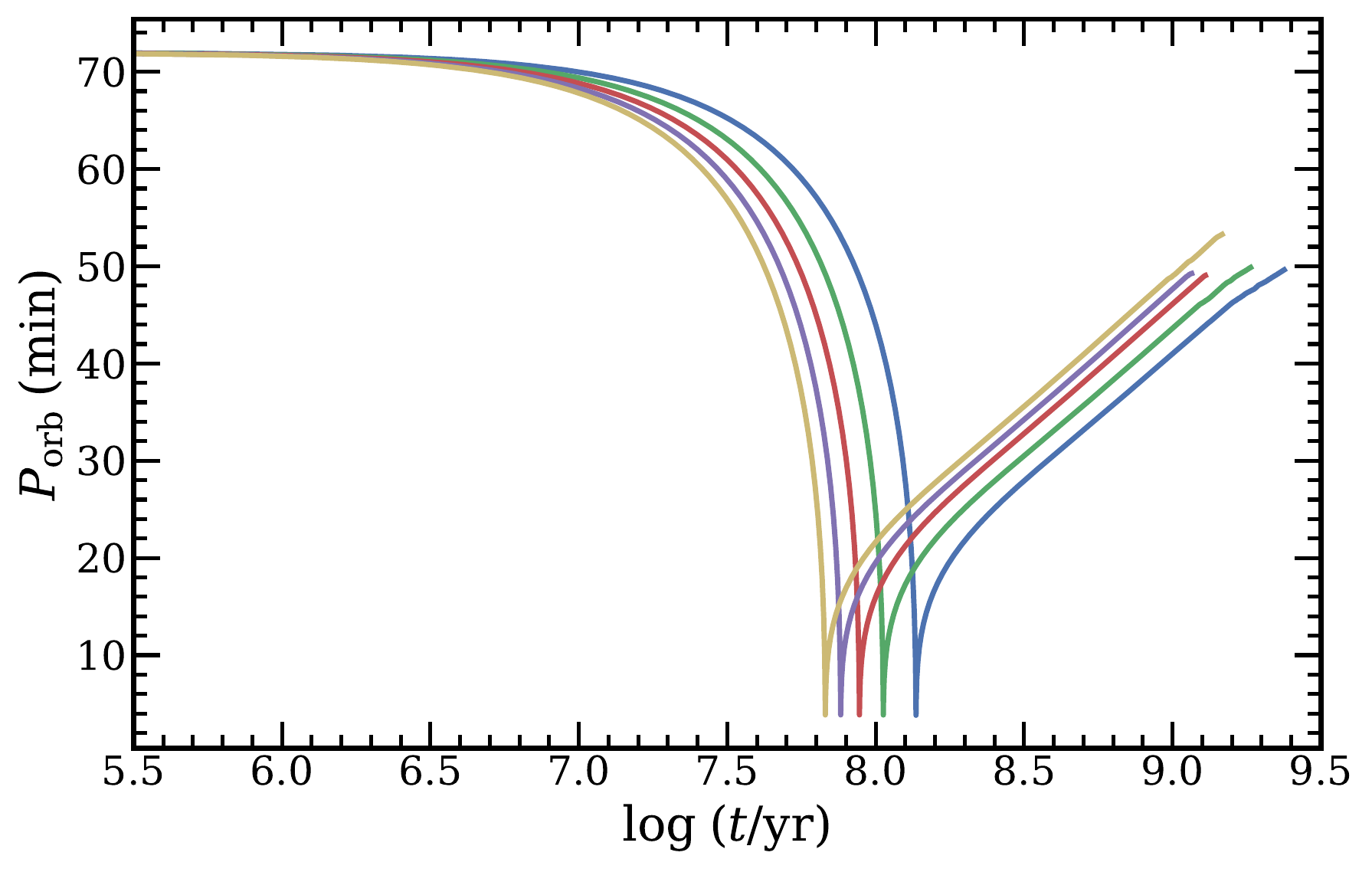}
    \includegraphics[width=\columnwidth]{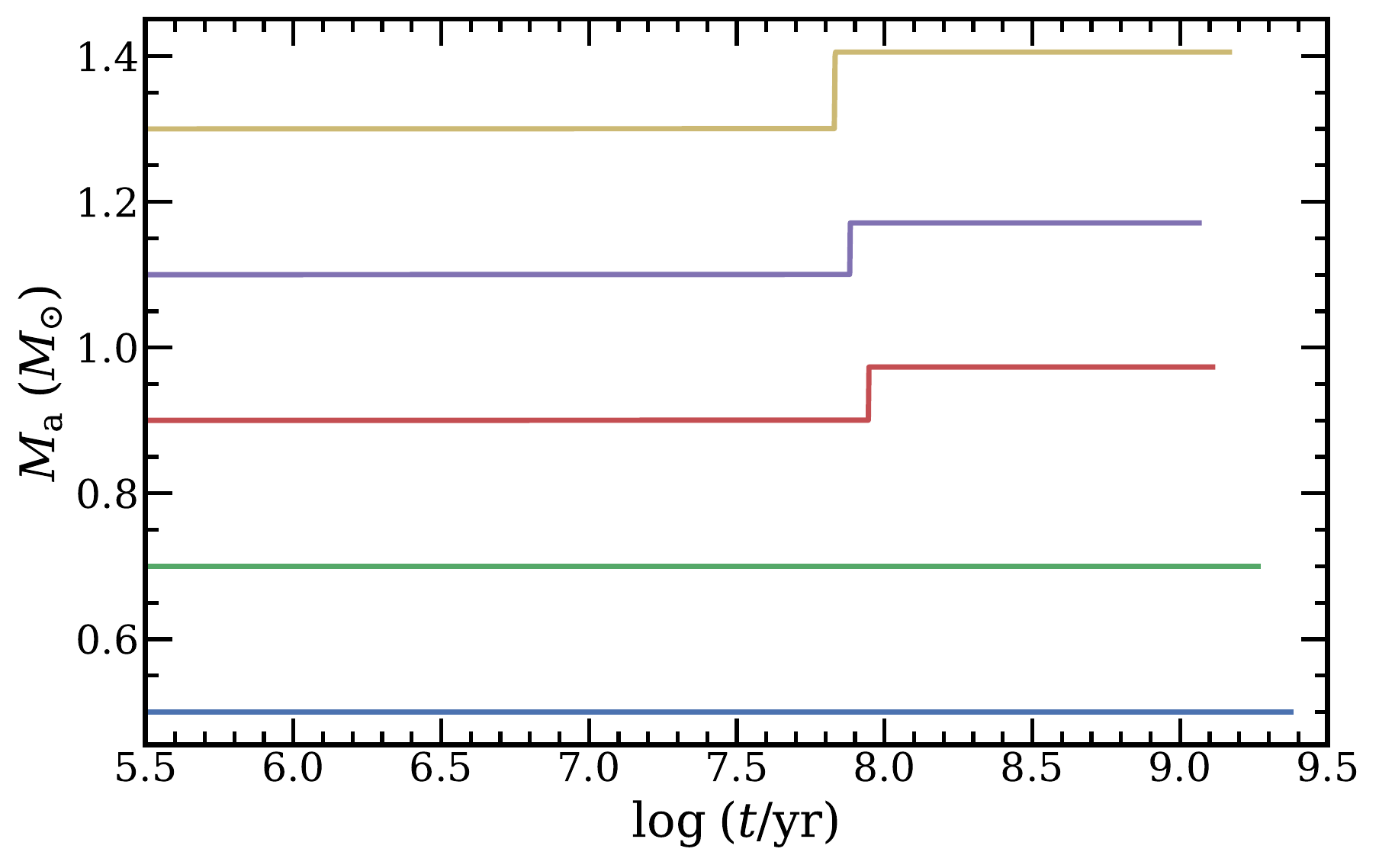}
    \caption{Evolution of mass transfer rate (upper panel), orbital period (middle panel) and accretor mass (bottom panel) for AM CV binaries with different accretor masses. In the upper panel, the arrows indicate the evolutionary direction. In these binaries, the initial donor masses and the initial orbital periods are the same, i.e. $0.21\;M_{\odot}$ and $0.05\;$days, respectively. }
    \label{fig:md_pd_lgmdot_diff_acc}
\end{figure}

\begin{figure}
    \centering
    \includegraphics[width=\columnwidth]{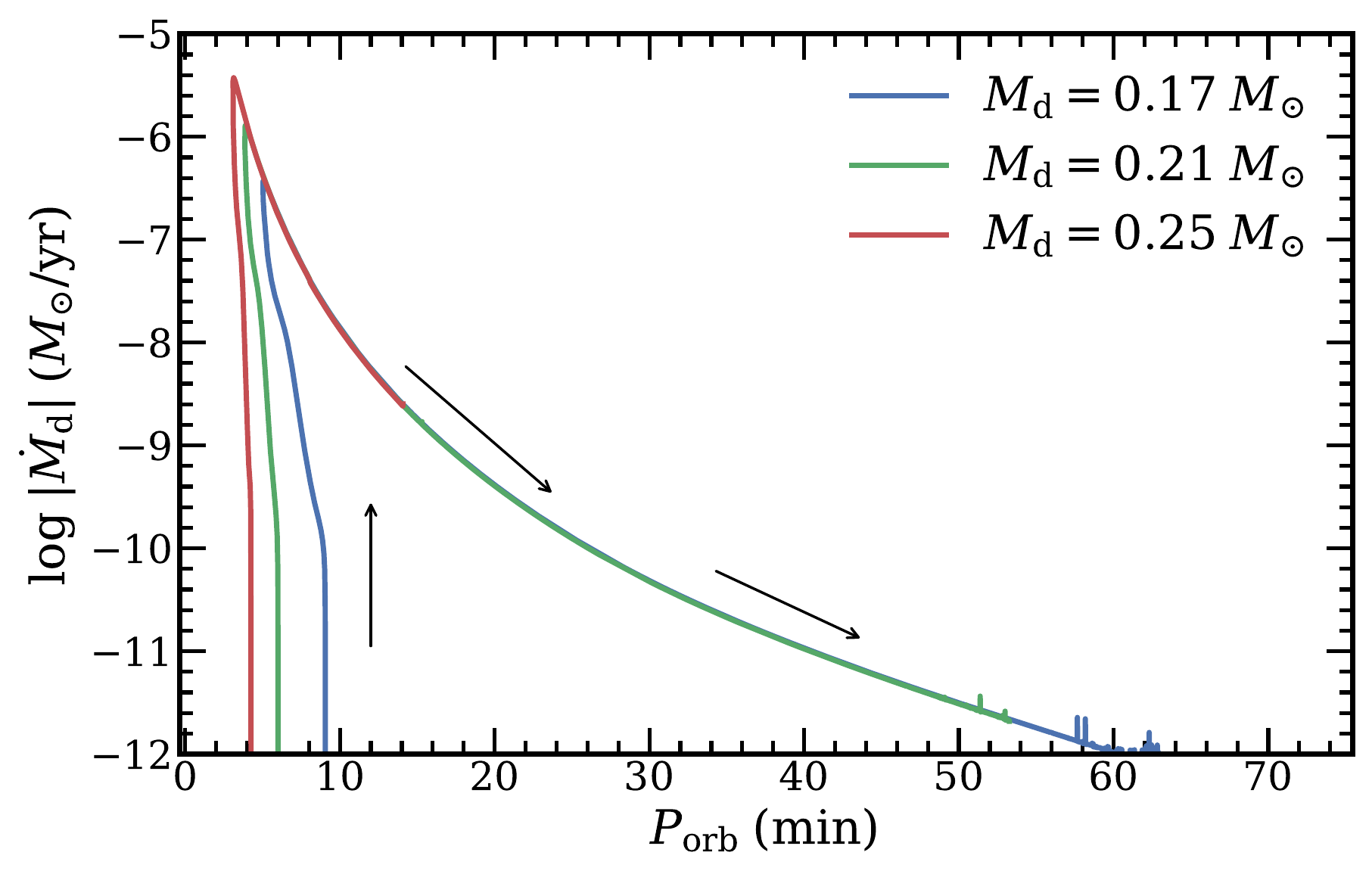}
    \includegraphics[width=\columnwidth]{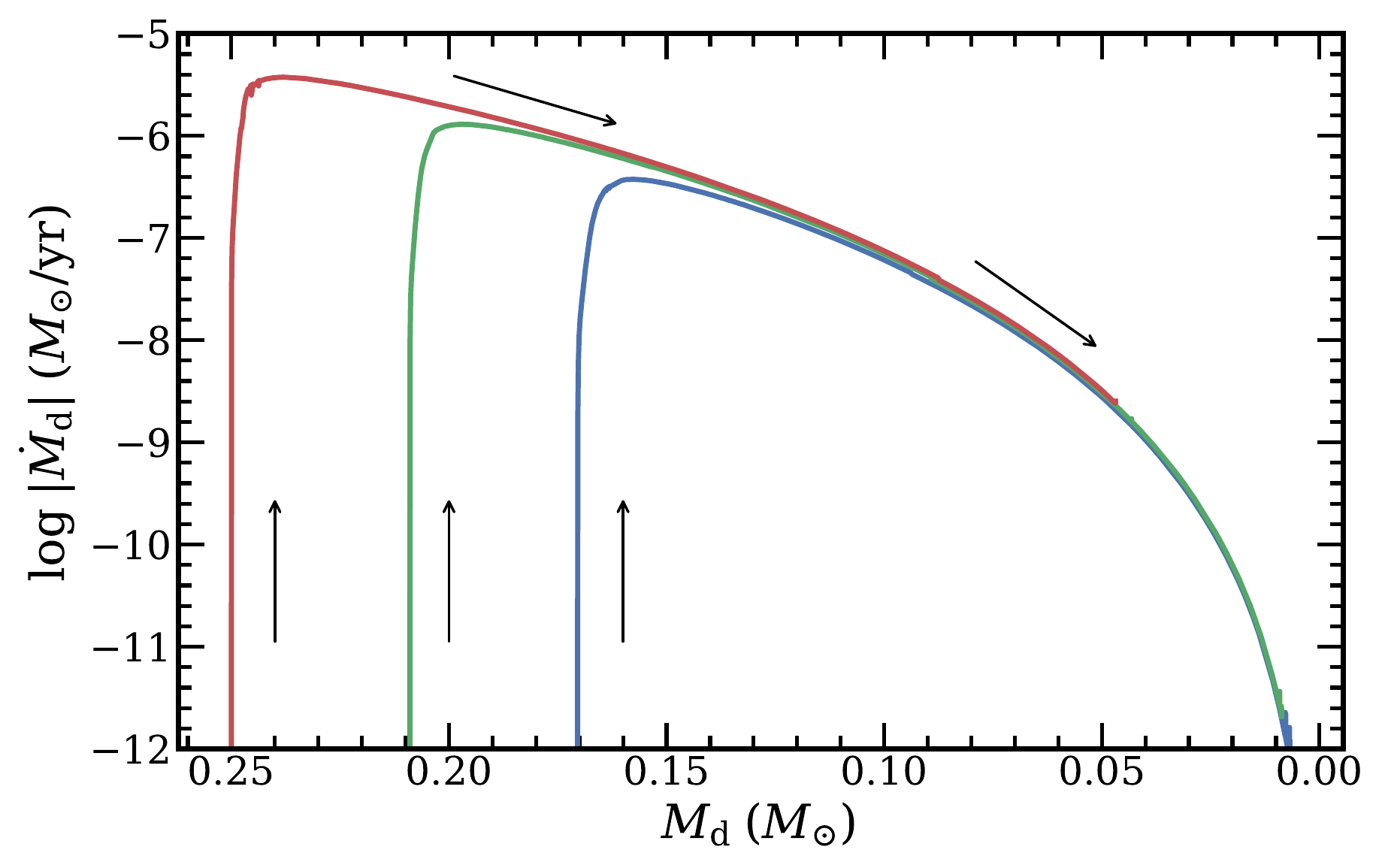}
    \caption{Evolution of mass transfer as a function of orbital period (upper panel) and He WD mass (lower panel). In the plot, the arrows indicate the evolutionary direction of binaries. 
    In these binaries, the initial masses of WD accretors and the initial orbital periods are the same, i.e. $1.30\;M_{\odot}$ and $0.05\;$days, respectively. In these plot, we do not show these evolutionary tracks for these binaries with donor masses $M_{\rm d} \ge 0.30\;M_{\odot}$. This is because their mass transfer rates during their evolution can exceed the Eddington limit and we assume these binaries will merge. }
    \label{fig:md_pd_lgmdot_diff_md}
\end{figure}

In Fig.~\ref{fig:md_pd_lgmdot_diff_acc}, we show the evolution of mass transfer rate, orbital period and accretor mass for binaries with different WD accretor masses and a same donor mass.  From the upper and middle panels, we can find the evolution of mass transfer rate and orbital period are very similar for these binaries. The minimum orbital periods during the evolution are almost the same for these binaries. This is because the evolution of orbital period during the mass transfer mainly depends on the donor mass \citep{ctch22b}. In the binary system with an accretor mass of $1.30\;M_{\odot}$, the accretor mass reaches $1.40\;M_{\odot}$ during its evolution.  The WD accretor can collapse into a NS and we do not stop the calculation at that point. We have a further discussion on this kind of systems in Sec.~\ref{sec:aic}. For the systems with initial accretor masses of $0.50$ and $0.70\;M_{\odot}$, the accretor masses do not change during their evolution. This is because we assume that the mass transfer is completely non-conservative for these systems with accretor masses $M_{\rm a} \le 0.75\;M_{\odot}$. 

Fig.~\ref{fig:md_pd_lgmdot_diff_md} presents the evolution of mass transfer rate as a function of orbital period and He WD masses for binaries with different He WD masses and a same accretor mass. From this plot, we can also find that these tracks converge to a single branch after the peaks of mass transfer rate.   
Compared with these systems with smaller He WD masses,  the systems with massive He WDs have larger maximum mass transfer rate and smaller minimum orbital period. This is mainly because massive He WDs have smaller radius.

\section{Discussion}
\label{sec:dis}

\subsection{Properties as GW sources}
\label{sec:dis_gw}

\begin{figure*}
    \centering
    \includegraphics[width=\columnwidth]{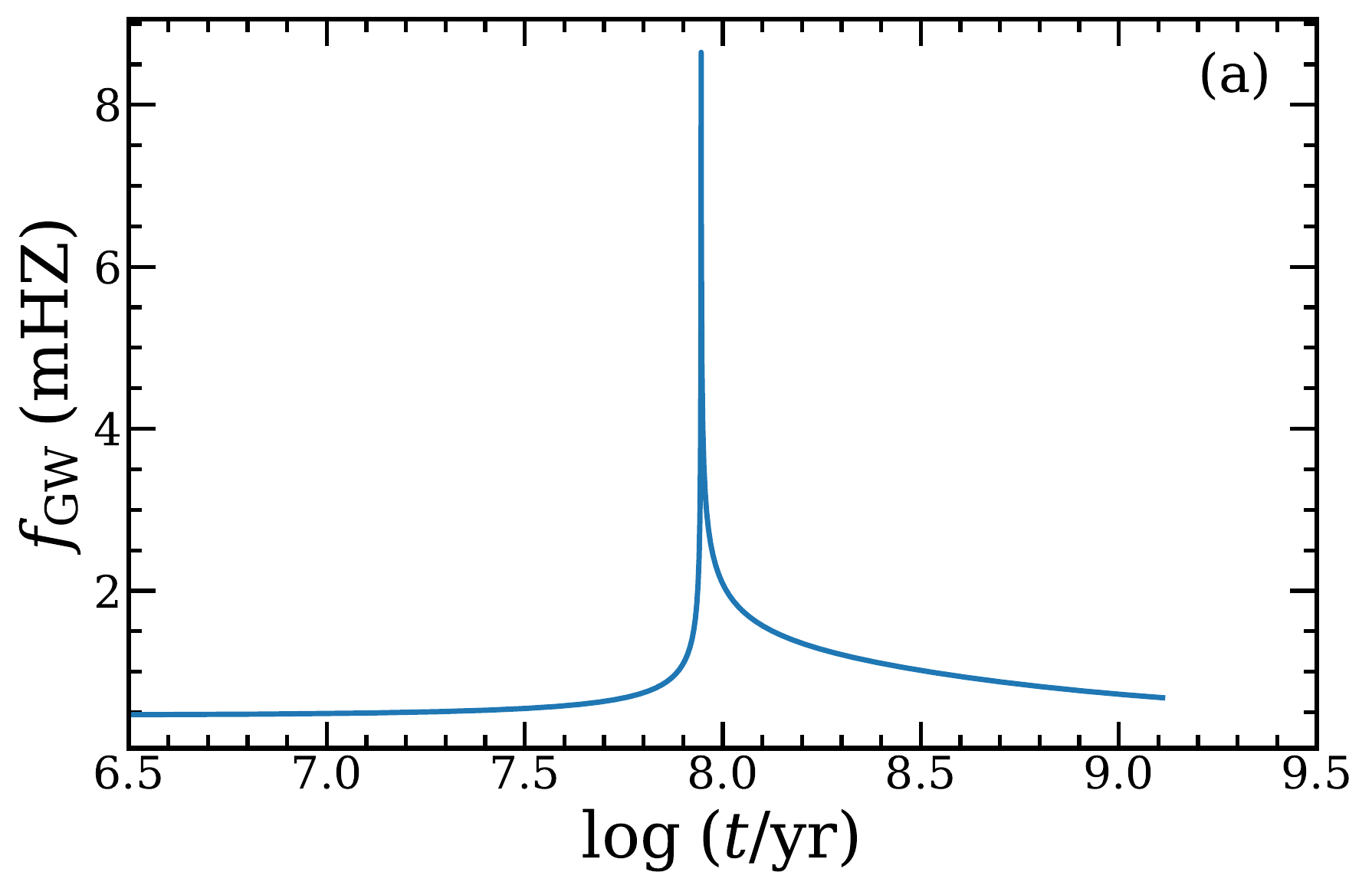}
    \includegraphics[width=\columnwidth]{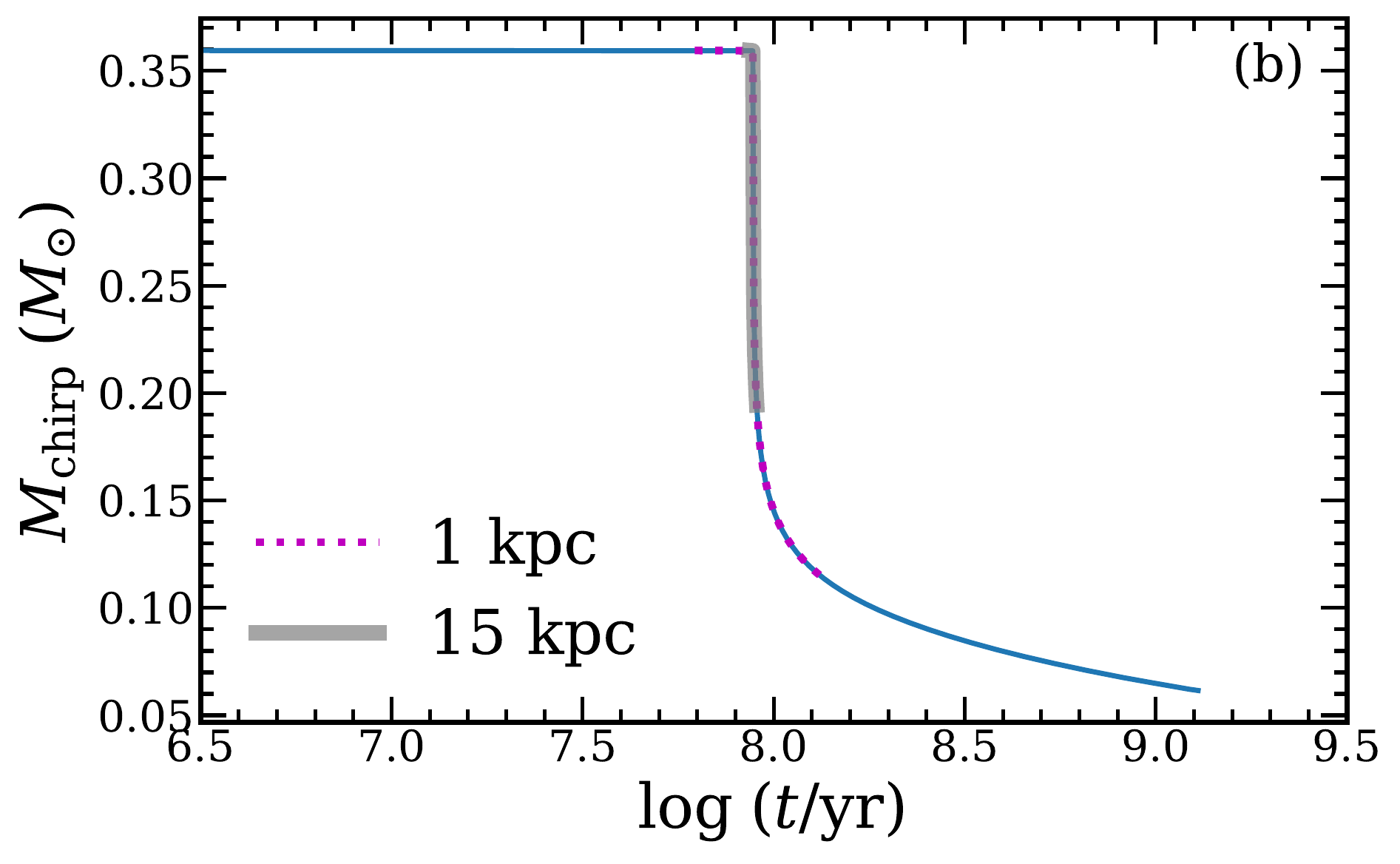}
    \includegraphics[width=\columnwidth]{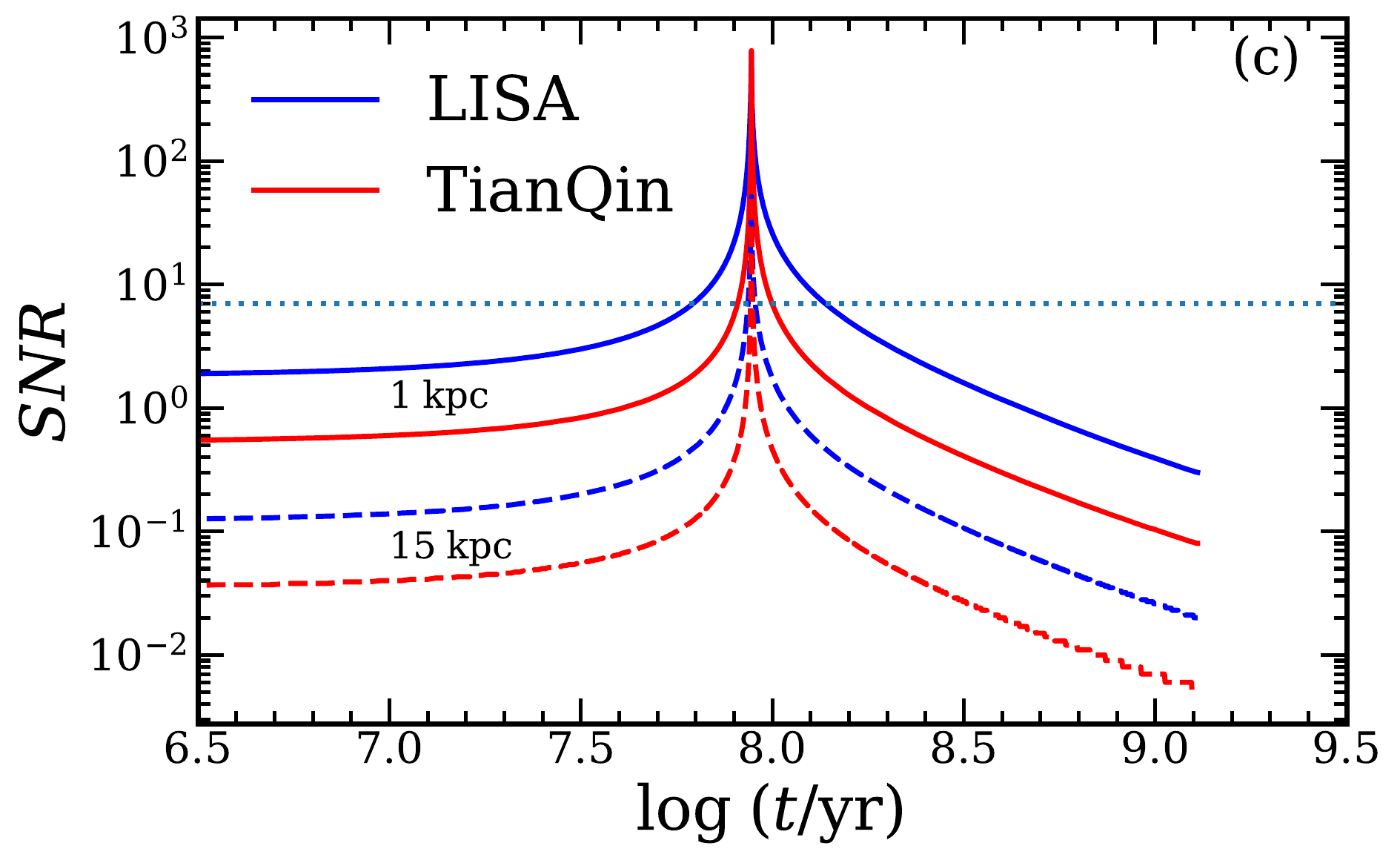}
    \includegraphics[width=\columnwidth]{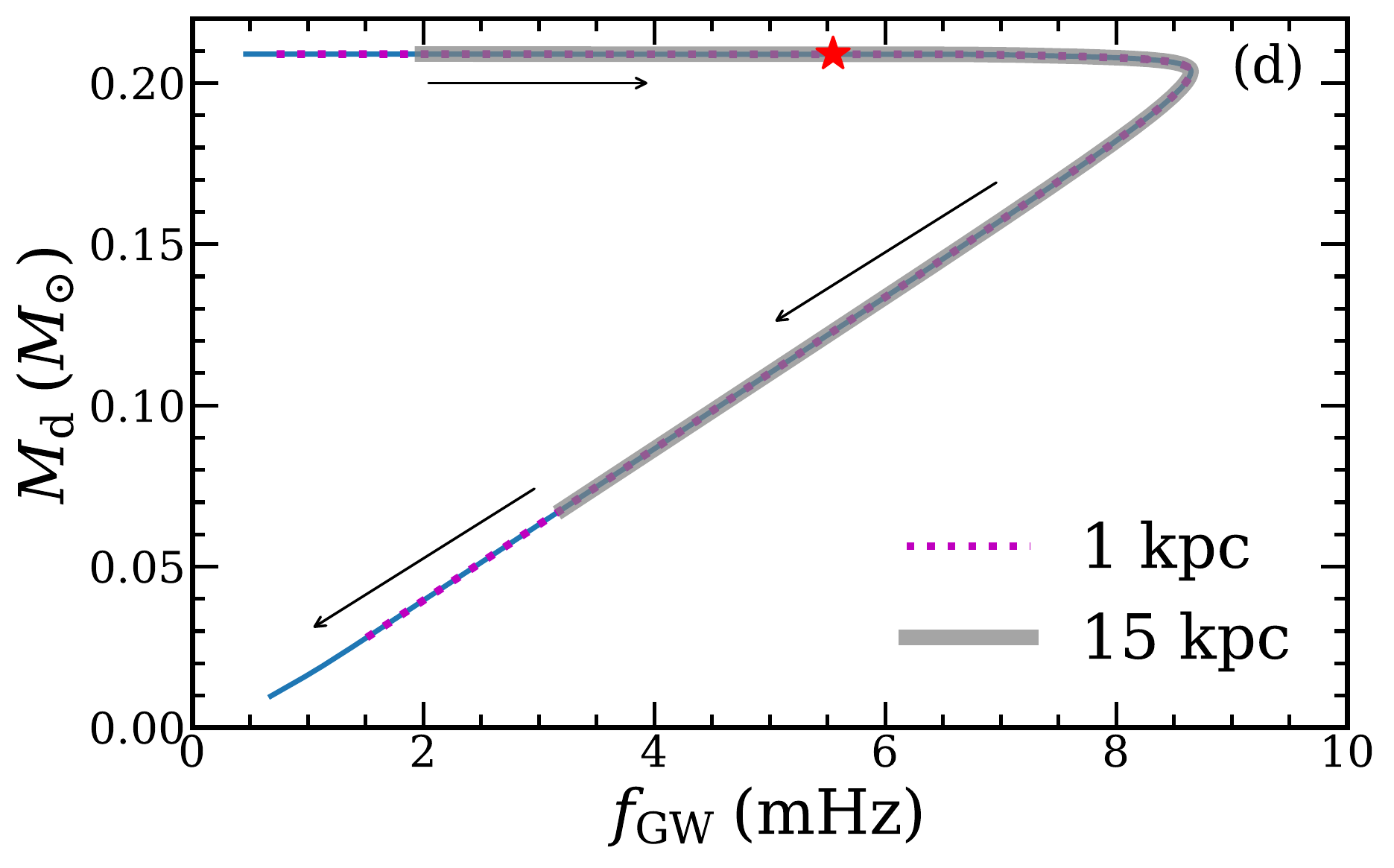}
    \caption{Evolution of GW frequency (panel a), chirp mass (panel b) and signal-to-noise ratio (panel c) as a function of time. Panel (d) shows the evolution of donor mass as a function of GW frequency. 
    In this example, the initial binary parameters are $M_{\rm a} = 0.90\;M_{\odot}$, $M_{\rm d} = 0.21\;M_{\odot}$ and $P_{\rm orb} = 0.05\;$days. In the panel (c), the blue and red lines are computed with the sensitivity curves of LISA with an observing time of $T = 4\;$yr and TianQin with an observing time of $T = 5\;$yr, respectively. The solid and dashed lines are for AM CVn binaries at a distance of $1\;$ and $15\;$kpc, respectively. The dotted line indicates the critical SNR $= 7$, above which the source becomes detectable. In the panel (b) and (d), the magenta dashed and grey solid colors indicate the ranges in which the source is detectable by LISA at a distance of $1\;$kpc and 15\;kpc, respectively. The star symbol indicates the onset of mass transfer (${\rm log}(\dot{M}/(M_{\odot}/{\rm yr})) >= -12.0$). }
    \label{fig:gw_sg}
\end{figure*}

\begin{figure}
    \centering
    \includegraphics[width=\columnwidth]{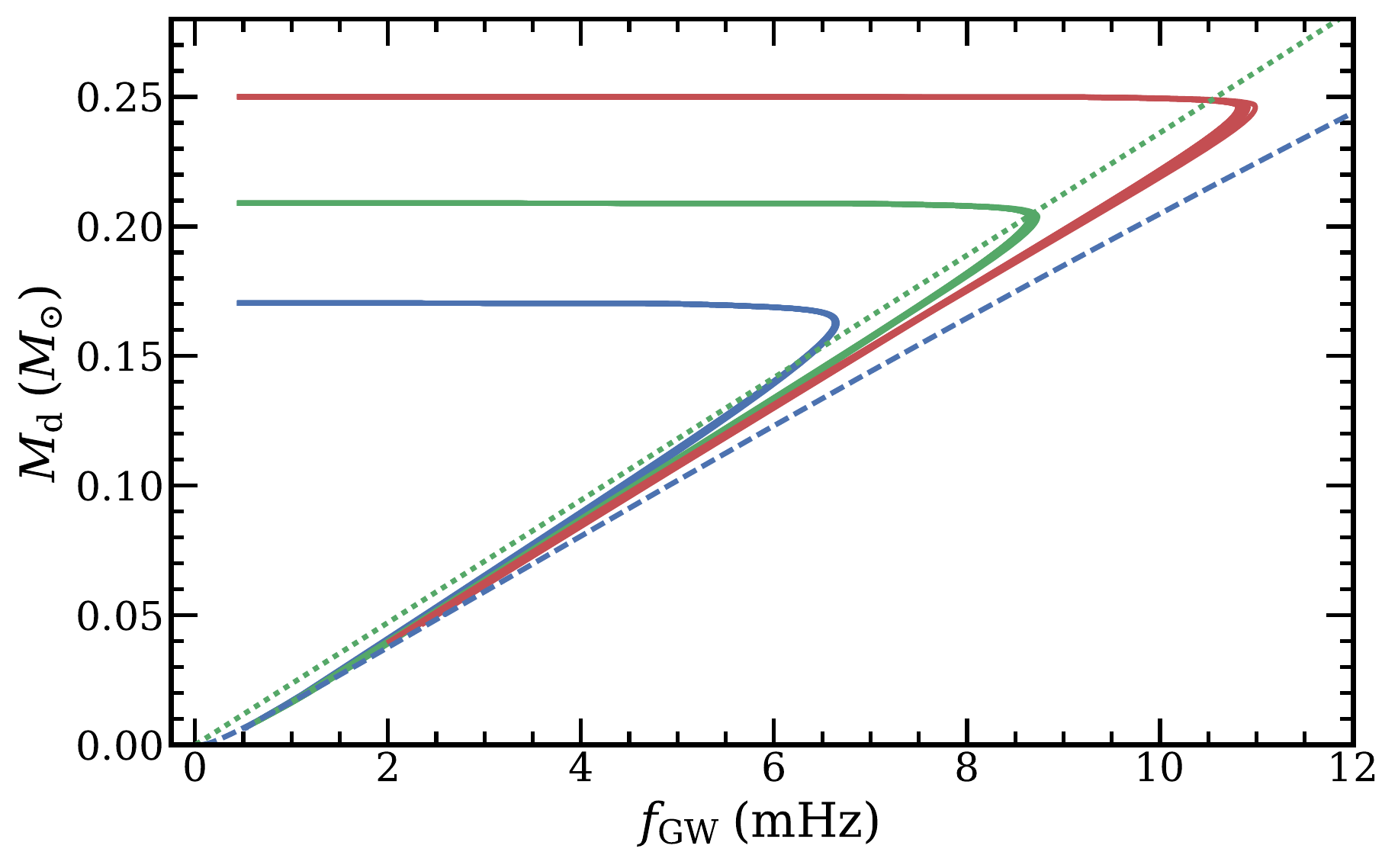}
    \caption{Evolution of donor mass as a function of GW frequency for all binary systems of which mass transfer rates do not exceed the Eddington mass transfer rate. In this plot, these binaries with a same donor mass have very similar evolutionary tracks. Therefore, it appears there are only three tracks in the plot. The dotted and dashed line indicate the analytic relations from \citet{ctch22b} (see their eqs. 7 and 13) .   }
    \label{fig:md_lgf}
\end{figure}

Given the short orbital periods of AM CVn binaries, they are expected to be important sources of low GW frequency. In Fig.~\ref{fig:gw_sg}, we present an example of the evolution of GW frequency, chirp mass and signal-to-noise ratio (SNR) as a function of time for a binary system. In addition, the evolution of donor mass as a function of GW frequency is also shown. 
Here the SNR for LISA and TianQin is computed with the Python package \textsc{LEGWORK} \citep{wbd22}. 
The initial binary parameters are $M_{\rm a} = 0.90\;M_{\odot}$, $M_{\rm d} = 0.21\;M_{\odot}$ and $P_{\rm orb} = 0.05\;$days.
From these plots, we can find that the AM CVn binaries have a strong GW emission in the mHZ regime. The SNR can be up to $\sim 800$ ($\sim 50$) if the source is loacted at 1\;kpc (15\;kpc). If we adopt the critical SNR $= 7$, above which the source becomes detectable, then we can find that this source can be detected by LISA and TianQin. The chirp mass for the detectable source is between $0.11\;M_{\odot}$ and $0.36\;M_{\odot}$ ($0.19\;M_{\odot}$ and $0.36\;M_{\odot}$) if the source is located at a distance of $1\;$kpc (15\;kpc). The donor mass for the detectable source is between $0.03\;M_{\odot}$ and $0.21\;M_{\odot}$ ($0.07\;M_{\odot}$ and $0.21\;M_{\odot}$) if the source is located at a distance of $1\;$kpc (15\;kpc). From panel (d), we can find there is a tight relation between the donor mass and GW frequency after the maximum frequency.

In Fig.~\ref{fig:md_lgf}, we present the evolution of WD donor mass as a function of GW freqency for all binaries. In the plot, we do not show these binaries with donor masses larger than $0.30\;M_{\odot}$. This is because the mass transfer rates of these systems exceed the Eddington limit and these systems are assumed to merge in our calculation. These systems with a same donor mass have almost the same evolutionary track in this plot. Therefore, it appears that there are only three lines in the plot. From this plot, we can find that all evolutionary tracks converge to the same branch after the maximum peak in frequencies are reached. In other words, there is a linear relation between the donor mass and GW frequency. This relation has been analytically derived by \citet{ctch22b} and can be described with their eqs. 7 and 13. A fitting formula for this relation has been proposed by \citet{bkbl+18} (see their fig. 1). With this relation, we can infer the He WD mass if the GW frequency is measured. Then the accretor mass can also be derived if the chirp mass is observed.

\subsection{Stability of mass transfer}
\label{sec:dis_mt}

\begin{figure}
    \centering
    \includegraphics[width=\columnwidth]{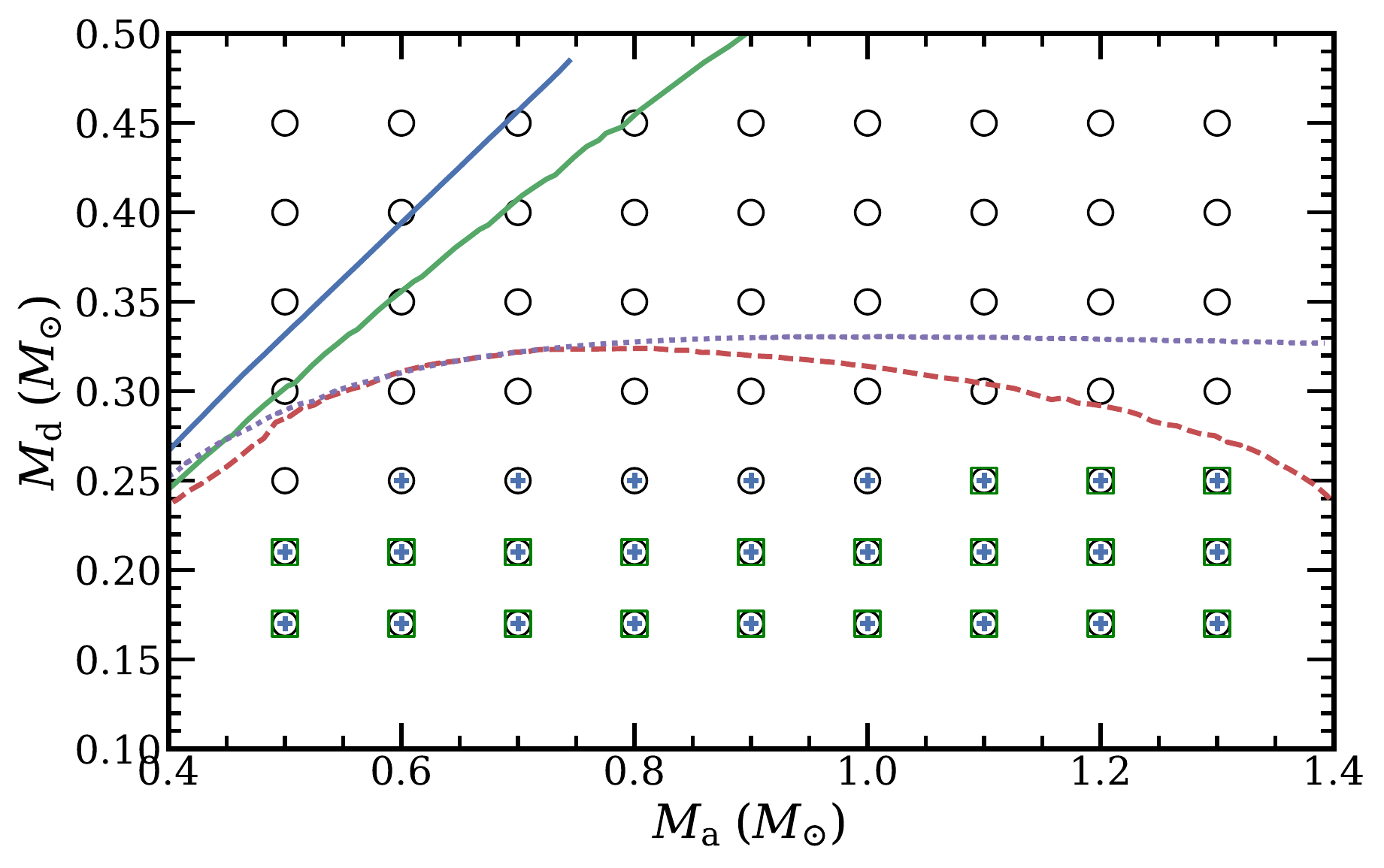}
    \caption{Stability limits for mass transfer in double WD binaries. The blue and green solid lines show the dynamical stability limits from \citet{npvy01} (see their eq. 4) and \citet{mns04} (see their eq. 30). The purple dotted and red dashed lines show the threshold of Eddington limit from \citet {hw99} and \citet{npvy01}, respectively. Binaries above these lines have mass transfer rates exceed the Eddington limit in their calculations.  The open circles show all the binaries we computed and all these binaries have dynamically stable mass transfer. The crosses indicate these binaries with maximum mass transfer rates smaller than the Eddington limit in our calculation. And these squares indicate these binaries with maximum mass transfer rates below the maximum stable burning rate of He. 
    }
    \label{fig:mt_sta}
\end{figure}

The stability of mass transfer of double WDs is still under debate. This problem could strongly influence the number of AM CVn and GW sources predicted by binary population synthesis model.

In Fig.~\ref{fig:mt_sta}, we present the stability limits for mass transfer of AM CVn binaries and compare our results with previous works.  
Following \citet{ch08}, if a binary system has a runaway mass transfer, we assume that the binary system will have dynamically unstable mass transfer. With this criteria, we can find that all binaries in our calculation have dynamically stable mass transfer. In addition, following \citet{mns04} and \citet{kblk17}, if we adopt a critical mass transfer rate of $\dot{M} = 0.01\;M_{\odot}/{\rm yr}$ as the limit for dynamically stable mass transfer, we can also find that all binaries in our calculation have dynamically stable mass transfer. 
This is very different from \citet{npvy01} and \citet{mns04} (see the solid lines in Fig.~\ref{fig:mt_sta}). The mass transfer of these binaries with smaller accretor and larger donor masses are dynamically stable in our calculation and unstable in \citet{npvy01} and \citet{mns04}. 
This is partially due to that we have non-conservative mass transfer in our calculation, while they assumed conservative mass transfer in their calculation (see sec.~\ref{sec:inf_eff} for detail discussion). 
In addition, a zero temperature is assumed for the WD donors in \citet{npvy01} and \citet{mns04}, which may lead to higher mass transfer rate (see sec.~\ref{sec:dis_pt} for more discussion). 

\citet{hw99} suggested that a common envelope may form if the mass transfer rate in double WD binaries is larger than the Eddington limit. With this restriction, we can find the threshold for Eddington-limited mass transfer and show these binaries below the threshold with crosses in Fig.~\ref{fig:mt_sta}. 
From the plot, we can find that the threshold in our calculation is slightly below that found by \citet{hw99} and \citet{nypv01}.  This is because the Eddington limit we adopted is lower than that in \citet{hw99} and \citet{nypv01}.

In our calculation, we assume that the optically thick wind occurs if the mass transfer rate is larger than the maximum stable burning rate of He. But there is another possibility in this regime. For example, \citet{pty14} found that the accreting WD may evolve into a red giant. In this scenario, the binary may enter common envelope and merge eventually. With this scenario in mind, we can find all binaries with maximum mass transfer rates smaller than the maximum stable burning rates of He \footnote{Since Eq.~\ref{eq:up} is only validated for WDs with masses $M_{\rm a} \ge 0.75\;M_{\odot}$, we adopt the maximum stable burning rate of He from \citet{pty14} for WDs with masses smaller than $0.75\;M_{\odot}$.}, which are shown as empty squares in Fig.~\ref{fig:mt_sta}. This may provide a new criterion for double WDs surviving mass transfer, which is slightly below the threshold for Eddington-limit.

\citet{cmsg+13} calculated the AM CVn space density to be $(5\pm 3) \times 10^{-7}\;{\rm pc}^{-3}$, which is 50 times smaller than the predicted value by the optimistic population synthesis model from the \citet{npvy01}. 
Recently, \citet{vkgw+22} found a space density of $6^{+6}_{-2} \times 10^{-7}\;{\rm pc}^{-3}$ and confirmed this discrepancy. 
Our new stability limits may have an important implication for the formation of AM CVn binaries in binary population synthesis study. 
Since the stability limits in our calculation are lower than that in \citet{npvy01}, we expect that less binaries will have stable mass transfer and less AM CVn binaries will be produced in the population synthesis models. This will be helpful to mitigate this discrepancy between the theoretical model and observation. 

It is also worth noting that we do not consider the rotation of the accretors in our calculation. During the mass transfer, the accretor may be spun up due to accretion. The coupling between the accretor's spin and orbit may lead to extra orbital angular momentum loss during binary evolution. \citet{mns04} showed that the coupling may have an important effect on the stability of mass transfer depending on the synchronization timescale. On the other hand, \citet{ksgm+16} have shown that the accretor velocities of two AM CVn systems, GP Com and V396 Hya, are much slower than the critical. 

\subsection{Accretion induced collapse of ONe WDs}
\label{sec:aic}

As we show in Fig.~\ref{fig:md_pd_lgmdot_diff_acc}, some ONe WDs in our calculation can increase their masses to the Chandrasekhar mass limit ($\sim 1.40\;M_{\odot}$) and will collapse into NSs. At this point, the He WD masses are around $0.09 - 0.15\;M_{\odot}$. After the WD collapses into a NS, the binary system will become eccentric because of the mass loss during collapse and the natal kick of NS \citep[e.g.][]{tsyl13}. 
Because of mass loss during the collapse, the Roche lobe radius of the donors increase. Then these systems may evolve into detached binaries consisting of NSs and extreme low mass He WDs. Due to the GW radiation, the He WDs in these systems may fill their Roche lobe again at some point. Then these binaries will evolve into ultra-compact X-ray binaries. 

\subsection{Influence of accretion efficiency }
\label{sec:inf_eff}

\begin{figure}
    \centering
    \includegraphics[width=\columnwidth]{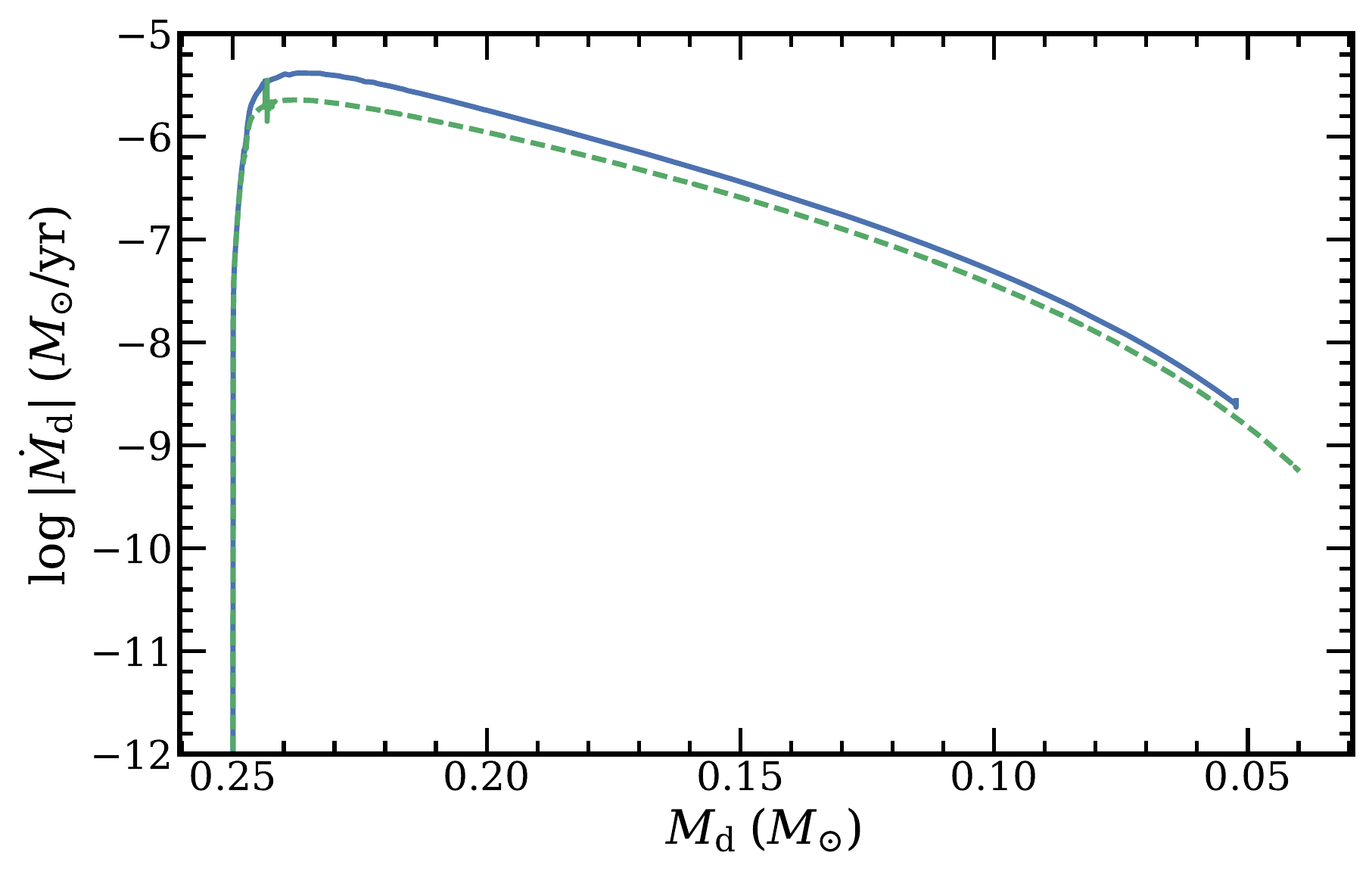}
    \caption{Comparison of the evolution of mass transfer rate between models with two different prescriptions of accretion efficiency. The blue solid (green dashed) line is for the model assuming conservative (completely non-conservative) mass transfer. The initial binary parameters in this example are $M_{\rm a} = 0.50\;M_{\odot}$, $M_{\rm d} = 0.25\;M_{\odot}$ and $P_{\rm orb} = 0.05\;$days. }
    \label{fig:com_diff_eff}
\end{figure}

In order to understand the influence of accretion efficiency, we compute the evolution of a binary system assuming the mass transfer is conservative, i.e. all the material lost by the donor is accreted by the accretor. 
In Fig.~\ref{fig:com_diff_eff}, we show the comparison of the evolution of mass transfer rate for binaries with different prescriptions of retention efficiency. From this plot, we can find that their evolution is very similar. But the mass transfer rate is relatively higher in the model assuming conservative mass transfer. Binaries with high mass transfer rates are more likely to be dynamically unstable. This may partially explain the difference of dynamical stability of mass transfer between our calculation and previous studies.

\subsection{Influence of initial effective temperature and orbital period}
\label{sec:dis_pt}

\begin{figure}
    \centering
    \includegraphics[width=\columnwidth]{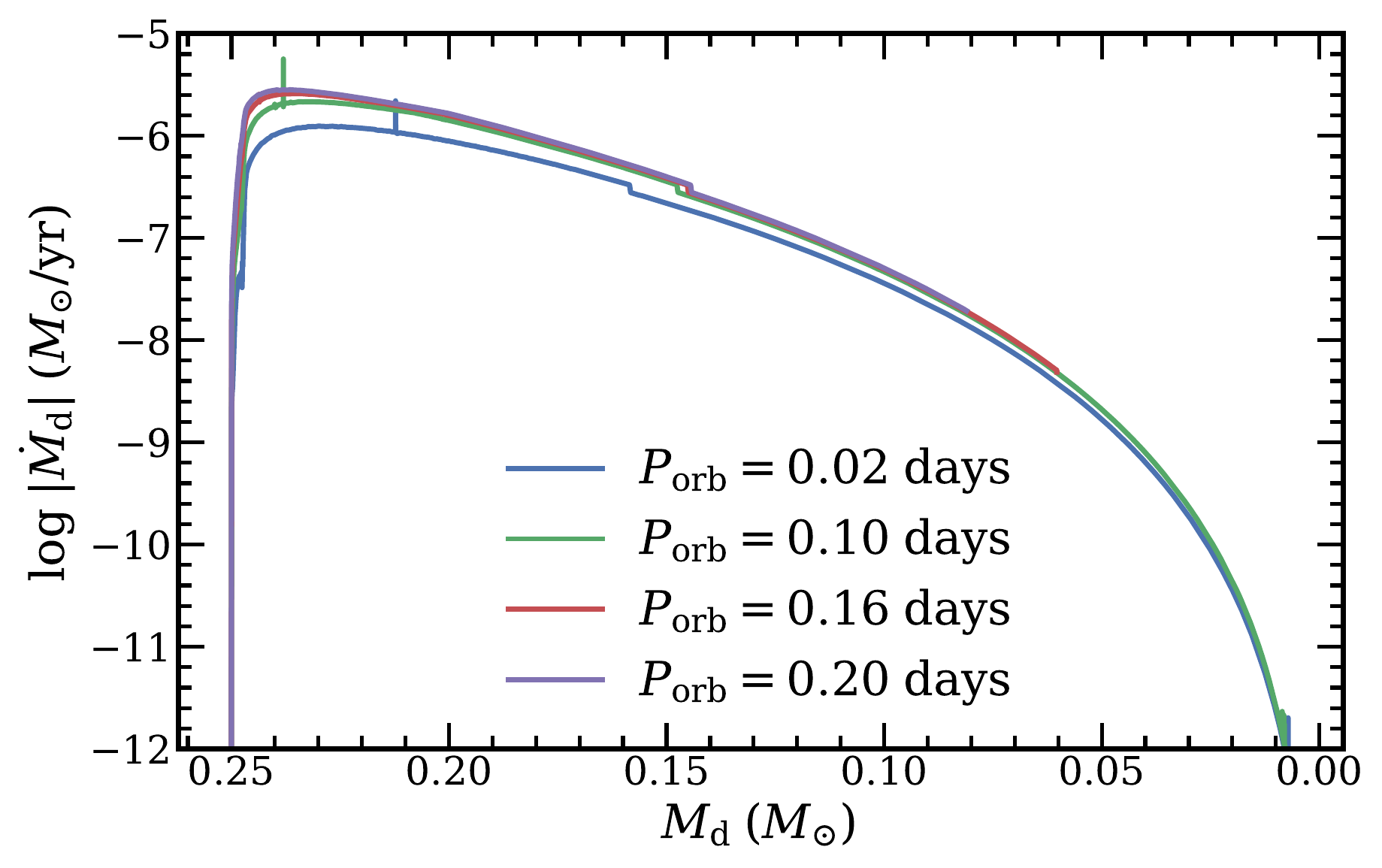}
    \caption{Evolution of mass transfer rate as a function of donor mass for binaries with different orbital periods. In these binaries, the initial WD masses are the same, i.e. $M_{\rm a} = 0.80\;M_{\odot}$ and $M_{\rm d} = 0.25\;M_{\odot}$. The initial orbital periods are $0.02$, $0.10$, $0.16$ and $0.20\;$days.}
    \label{fig:com_diff_porb}
\end{figure}

In our calculation, the initial temperatures of He WDs and orbital periods are assumed to be the same for all binaries. This may be not realistic, given that the double WD systems can be produced from stable mass transfer and common envelope ejection channels \citep{lcch19}. The He WDs in these binaries with long (short) orbital periods have long (short) time to cool down, leading to low (high) temperatures at the onset of mass transfer. Therefore, the influence of initial effective temperature and orbital period should be similar. 

In order to understand their influence, we make a new $0.25\;M_{\odot}$ He WD model with a high temperature following the method described in Sec.~\ref{sec:ini_wd}. The central and effective temperatures in this model are $T_{\rm c} = 3.24 \times 10^{7}\;$K and $T_{\rm eff} = 2.74\;\times 10^{4}\;$K, respectively. With this model, we compare the evolution of mass transfer rate for binaries with different initial orbital periods, which is shown in Fig.~\ref{fig:com_diff_porb}. From this plot, we can find that the mass transfer rate is slightly higher for binaries with larger orbital periods at early phase of mass transfer. At later times, the difference is very small.

\section{Conclusions}
\label{sec:con}

In this work, we have comprehensively studied the evolution of AM CVn binaries with WD donors using the stellar evolution code \textsc{mesa}.  Instead of simply assuming conservative mass transfer, we have considered the dependence of retention efficiency on accretor mass and accretion rate. In our calculation, the accretor mass ranges from $0.50$ to $1.30\;M_{\odot}$ and the donor mass ranges from $0.17$ to $0.45\;M_{\odot}$. The main results are as follows:

\begin{itemize}
    \item These binaries with same He WD masses but different accretor masses  have very similar evolution and similar minimum orbital periods. These binaries with same accretor masses but larger donor masses have larger maximum mass transfer rate and smaller minimum orbital periods.
    
    \item We demonstrate that the GW signal from AM CVn binaires can be deteced by LISA and TianQin. Moreover, there is a linear relation between the WD donor mass and GW frequency during the mass transfer phase.
    
    \item In our calculation, all binaries have dynamically stable mass transfer, which is very different from previous studies. The threshold donor mass for Eddington-limited mass transfer for a given accretor mass is lower than previous studies. Assuming that the binary may enter common envelope if the mass transfer rate is larger than the maximum stable burning rate of He, we provide a new criterion for double WDs surviving mass transfer, which is slightly below the threshold of Eddington limit (see Fig.~\ref{fig:mt_sta}). 
    
    \item These binaries with ONe WD accretors may evolve into binaries consisting of NS and extremely low mass WDs and further ultra-compact X-ray binaries.

\end{itemize}

\begin{acknowledgments}
HLC thanks Matthias Kruckow for useful comments. 
This work is partially supported by the National Key R\&D Program of China (grant Nos. 2021YFA1600403, 2021YFA1600401), the National Natural Science Foundation of China (Grant No. 12090040/12090043, 12073071, 11873016, 11733008),
Yunnan Fundamental Research Projects (Grant No. 202001AT070058, 202101AW070003), the science research grants from the China Manned Space Project with No. CMS-CSST-2021-A10. The authors gratefully acknowledge the ''PHOENIX Supercomputing Platform'' jointly operated by the Binary Population Synthesis Group and the Stellar Astrophysics Group at Yunnan Observatories, CAS.
We are grateful to the \textsc{mesa} council for the \textsc{mesa} instrument papers and website. 
\end{acknowledgments}

\vspace{5mm}
\software{MESA \citep{pbdh+11,pcab+13,pmsb+15,psbb+18,pssg+19}, LEGWORK \citep{wbd22}, matplotlib \citep{hunt07}, numpy \citep{numpy20}, astropy \citep{astropy13,astropy18},
Python from \dataset[python.org]{https://www.python.org/}}.

\bibliography{hailiang_refs}{}
\bibliographystyle{aasjournal}

\end{document}